\documentclass[12pt]{article}

\usepackage[a4paper,left=30mm,right=20mm,top=20mm,bottom=20mm]{geometry}
\usepackage{graphics}
\usepackage{amsmath}
\usepackage{epsfig}
\usepackage{cite}
\usepackage{xcolor}
\usepackage[unicode]{hyperref}
\title{Testing the nuclear TMD gluon densities with heavy flavor production in proton-lead collisions at LHC}

\author{A.V.~Lipatov$^{1}$, A.V.~Kotikov$^{2}$}

\begin{document}

\maketitle

\begin{center}
{\it $^{1}$Skobeltsyn Institute of Nuclear Physics, Lomonosov Moscow State University, 119991 Moscow, Russia\\}
{\it $^{2}$Joint Institute for Nuclear Research, 141980 Dubna, Moscow region, Russia}\\

\end{center}

\vspace{0.5cm}

\begin{center}

{\bf Abstract }

\end{center}

We employ a simple model for nuclear modification of ordinary parton densities in a proton
to evaluate the Transverse Momentum Dependent gluon and quark distributions in nuclei (nTMDs)
within the popular Kimber-Martin-Ryskin/Watt-Martin-Ryskin approach.
The model is based on a global analysis of available deep inelastic scattering data for
different nuclear targets within the rescaling model, incorporating Fermi motion effects.
The derived nTMDs
are tested with latest CMS data on inclusive $b$-jet and $B^+$ meson
production in proton-lead collisions collected at $\sqrt s = 5.02$ and $8.16$~TeV
using the High Energy Factorization framework.
We predict the corresponding nuclear medium modification factors 
to be about of
$0.8 - 1.2$ in the probed kinematical region,
which is consistent with other estimations.
Specially we highlight a possibility to investigate the
nuclear modification of parton densities by applying
different cuts on the final states
in such processes.

\indent

\vspace{1.0cm}


\vspace{1.0cm}

\newpage


It is known that, at present, any theoretical investigation of QCD processes 
is usually relies on a factorization of the effects of short and long distances at some scale.
A key ingredient of such consideration is the parton (gluon and quark) distribution functions in a proton (PDFs), $f_a(x,\mu^2)$
with $a = g$ or $q$ and $x$ being the longitudinal momentum fraction of proton carried by parton $a$,
which describe the parton content of a proton at the scale $\mu^2$. 
In the conventional QCD factorization, the PDFs obey the Dokshitzer-Gribov-Lipatov-Altarelli-Parisi (DGLAP) equations\cite{DGLAP}.
At high energies, an alternative description can be achieved within the High Energy Factorization\cite{HighEnergyFactorization}, or $k_T$-factorization approach\cite{kt-factorization},
which is based on the Balitsky-Fadin-Kuraev-Lipatov (BFKL)\cite{BFKL} or
Catani-Ciafaloni-Fiorani-Marchesini (CCFM)\cite{CCFM}
evolution equations for transverse momentum dependent (TMD, or unintegrated) gluon densities in a proton, $f_g(x,{\mathbf k}_T^2,\mu^2)$.
Such an approach has certain technical advantages in the ease of including higher-order
radiative corrections corresponding to the gluon emissions in the initial state (see, for example, reviews\cite{TMD-review-our, TMD-review} for more information).

When it comes to nuclei, a number of complications arise. As is known, the nucleus is far from a simple picture of quasi-free nucleons\cite{EMCEffect, NuclReview-1, NuclReview-2, NuclReview-3}. Instead one finds so-called nuclear shadowing, anti-shadowing, EMC effect and
Fermi motion dominance regions at $x \leq 0.1$, $0.1 \leq x \leq 0.3$, $0.3 \leq x \leq 0.7$ and $x \geq 0.7$, respectively. Defining the nuclear medium modification factor as a ratio of
per-nucleon deep inelastic structure functions in nuclei $A$ and deuteron
$R_{AD} = F_2^A(x,Q^2)/F_2^D(x,Q^2)$, one then can refer to
the shadowing and anti-shadowing effects as having $R_{AD} < 1$ and $R_{AD} > 1$,
whereas EMC effect and Fermi motion
correspond to a drop of $R_{AD}$ in the valence-dominant region and an
increasing of $R_{AD}$ at larger $x$. So, corresponding pQCD calculations demand nuclear PDFs and/or TMDs (nPDFs or nTMDs),
which essentially differ from the usual parton distributions in a proton.
Of course, their detailed knowledge  --- in particular, knowledge of nuclear gluon densities --- is necessary for any
theoretical description and proper interpretation of $pA$ processes studied at modern (LHC, RHIC) and future colliders (FCC-he, EiC, EicC, NICA).
In turn, $pA$ collisions is a natural reference for more complex nucleus-nucleus ($AA$) interactions,
where nuclear matter can reach extremely high energy densities and temperatures, transforming
into its hot and dense deconfined phase, the quark-gluon plasma (QGP)\cite{QGP-1, QGP-2}.

Despite significant theoretical efforts made in recent years,
nPDFs and espesially nTMDs are still poorly known at present
and have large uncertainties due to shortage of experimental data and/or
their limited kinematic coverage (see\cite{nPDFs-unc-1, nPDFs-unc-2, nPDFs-unc-3} and references therein).
Several approaches could be used to evaluate
these objects. In fact,
they could be extracted from a global fit to
nuclear data and
then their scale dependence can be determined as a solution of QCD evolution equations (see, for example,\cite{EPPS21, nNNPDF3, nCTEQ, nIMP}
for nPDFs and\cite{nTMDs-PB} for nTMDs).
This technique is in a close analogy with the standard derivation of the PDFs and/or TMDs.
The second strategy is based on special models\cite{KulaginPetti-1, KulaginPetti-2, RescalingModel-1, RescalingModel-2, GeometryScaling-1, GeometryScaling-2},
in particular, rescaling model\cite{RescalingModel-1, RescalingModel-2} and approaches where the
geometrical scaling\cite{GeometryScaling-1, GeometryScaling-2} is applied\cite{nGBW, nLLM}.
So, in the rescaling model it is assumed that the effective size of gluon and quark confinement
in the nucleus is greater than in the free nucleon, thus giving a shift in the hard scale $\mu^2$ (see also review\cite{RescalingModel-Review}).

In this note, we concentrate mainly on the nTMD gluon densities
and derive them from nPDFs using the popular Kimber-Martin-Ryskin (KMR)\cite{KMR-LO} or Watt-Martin-Ryskin (WMR)\cite{WMR-LO} approach.
The WMR prescription is an extension and further development of KMR formalism and
explored currently at the next-to-leading order (NLO)\cite{WMR-NLO}.
In the KMR/WMR method, the DGLAP strong ordering condition is relaxed at the last step of parton evolution cascade,
so the transverse momentum is no longer negligible compared to the evolution scale $\mu^2$,
thus giving rise to the transverse momentum dependence of the parton densities.
Such prescription, where ordinary PDFs are employed as an input for the KMR/WMR procedure,
is widely used in phenomenological applications (see, for example,\cite{Szczurek-HFjj,Szczurek-phi,Motyka-photon,Iran-WZ,KMR-VFNS-HF1,KMR-VFNS-HF2} and references therein).
The KMR/WMR method to generate nTMDs has been used already\cite{MRW-nucl, nTMDs-KMR}.
So, the analytical leading order (LO) expressions\cite{PDFs-DAS-1, PDFs-our-previous, RescalingModel-Apps1} (see also \cite{TMD-DAS1, TMD-DAS2}) for PDFs obtained at small $x$
in the generalized double asymptotic scaling approximation\cite{gDAS1, gDAS2, gDAS3}
have been extended to nPDFs using the rescaling model adopted for low $x$\cite{RescalingModel-Apps1} and then
applied as an input for the KMR/WMR formulas\cite{nTMDs-KMR}.

Recently, an early derivation\cite{PDFs-DAS-1, PDFs-our-previous, RescalingModel-Apps1}
was singificantly improved by taking into account the
exact asymptotics at low and large $x$
and by incorporating subasymptotic terms fixed by the momentum conservation and/or Gross-Llewellyn-Smith and Gottfried sum rules\cite{PDFsSumRules1, PDFsSumRules2}.
Moreover, a simple
model for nuclear modification of these improved
PDFs has been proposed very recently\cite{RescalingModel-Fermi}.
The approach\cite{RescalingModel-Fermi} is based on the global analysis of available deep inelastic scattering data for different
nuclear targets within the low-$x$ enchanced rescaling model combined with the effects of Fermi motion.
Here we employ the KMR/WMR formalism to calculate nTMDs within the proposed model\cite{RescalingModel-Fermi}.
Then, we will test the derived TMDs with inclusive beauty production in $pA$ collisions at the LHC.
Such processes are known to be sensitive to the gluon content of the nucleus and provide us with a
possibility to reconstruct the full map of the latter (see, for example,\cite{MapForTMD}).
Moreover, they provide a unique opportunity to study the QGP\cite{QGP-HF1, QGP-HF2}.
In fact, by studying how the nuclear medium affects heavy quarks, the
properties of the medium can be determined.
Using the Monte-Carlo event generator \textsc{pegasus}\cite{PEGASUS},
which implements now the TMD gluon dynamics in nuclei,
we
calculate the transverse momentum and rapidity distributions of $b$-jets and $B^+$ mesons
and compare our predictions with latest experimental data\cite{CMS-bjets, CMS-B, CMS-B8} collected by the
CMS Collaboration at $\sqrt s = 5.02$ and $8.16$~TeV.
Of course, these data are of a great
importance to test the derived predictions for nTMDs.
Then we investigate corresponding effects of nuclear medium modifications
in different kinematical regions.
The consideration below extends and continues the line of our previous studies\cite{RescalingModel-Apps1, RescalingModel-Fermi, nTMDs-KMR}.

We start from a very brief description of our approach.
In the fixed-flavor-number-sheme (FFNS) with $N_f = 4$, where $b$ and $t$ quarks are separated out,
the non-singlet ($NS$) and valence ($V$) parts of quark distribution functions at the LO can be represented in the following form\cite{PDFsSumRules2}:
\begin{gather}
	f_i(x,\mu^2) = \left[A_i(s)x^{\lambda_i}(1 -x) + \frac{B_i(s)\, x}{\Gamma(1+\nu_i(s))} + D_i(s)x (1 -x) \right] (1-x)^{\nu_i(s)},
	\label{eq1-NSV}
\end{gather}
\noindent
where $i = NS$ or $V$, $s = \ln \left[\alpha_s(\mu_0^2)/\alpha_s(\mu^2)\right]$ and
\begin{gather}
	A_{i}(s)=A_{i}(0) e^{-d(n_i)s}, \quad B_{i}(s) = B_{i}(0) e^{-p s}, \quad \nu_{i}(s)=\nu_{i}(0)+r s, \nonumber \\
	r = \frac{16}{3\beta_0}, \quad p=r\left(\gamma_{\rm E}+\hat{c}\right), \quad \hat{c}=-\frac{3}{4}, \quad d(n)=\frac{\gamma_{NS}(n)}{2\beta_0}, \quad n_i = 1 - \lambda_i.
	\label{eq2-NSV}
\end{gather}
\noindent
Here $\Gamma(z)$ is the Riemann's $\Gamma$-function, $\gamma_{\rm E} \simeq 0.5772$ is the Euler's constant,
$\beta_0 = 11 - 2N_f/3$ is the LO QCD $\beta$-function, $\lambda_{NS} = \lambda_V = 0.5$, $\gamma_{NS}(n)$ is the LO non-singlet anomalous
dimension and $A_i(0)$, $B_i(0)$ and $\nu_i(0)$ are the free parameters.
The term proportional to $D_i(s)$ is a subasymptotics one and its scale dependence is
fixed by the Gross-Llewellyn-Smith and Gottfreed sum rules\cite{PDFsSumRules2}:
\begin{gather}
	D_i(s) = \left(2 + \nu_i(s) \right) \left[N_i - A_i(s) {\Gamma(\lambda_i) \Gamma(2 + \nu_i(s)) \over \Gamma(\lambda_i + 2 + \nu_i(s))} - {B_i(s) \over \Gamma(2 + \nu_i(s))} \right],
	\label{eq3-NSV}
\end{gather}
\noindent
where $N_V = 3$ and $N_{NS} \equiv I_G(\mu^2) = 0.705 \pm 0.078$ (see\cite{PDFsSumRules2} and references therein for more information).
The singlet ($SI$) part of quark densities and gluon distribution
in a proton can be represented as combinations of "$\pm$" terms (see also\cite{PDFs-DAS-1, PDFs-our-previous, RescalingModel-Apps1}):
\begin{gather}
	f_i(x,\mu^2) = f_i^+(x,\mu^2) + f_i^-(x,\mu^2),
	\label{eq1-Sg}
\end{gather}
\noindent
where $i = SI$ or $g$ and
\begin{gather}
	f_{SI}^+(x,\mu^2) = \Bigg[ {N_f \over 9} \left(A_{g} + {4\over 9}A_q\right) \rho I_1(\sigma) e^{- \bar d^{+} s} (1-x)^{m_{q}^+} + D^+(s) \sqrt x (1 - x)^{n^+} - \nonumber \\
	- {K^+ \over \Gamma(2 + \nu^+(s))} \times {B^+(s)x \over \hat c - \ln(1 - x) + \Psi(2 + \nu^+(s))} \Bigg] (1 - x)^{\nu^{+}(s) + 1}, \label{eq-fSp} \\
	f_{SI}^-(x,\mu^2) = \Bigg[ A_{q}e^{- d^{-} s} (1-x)^{m_{q}^-} + {B^-(s)x \over \Gamma(1 + \nu^-(s))} + \nonumber \\
	+ D^{-}(s) \sqrt x (1 -x)^{n^-} \Bigg] (1 - x)^{\nu^{-}(s)}, \label{eq-fSm} \\
	f_{g}^+(x,\mu^2) = \Bigg[ \left(A_{g} + {4\over 9}A_q\right) I_0(\sigma) e^{- \bar d^{+} s}(1 - x)^{m_{g}^+} + {B^+(s) x\over \Gamma(1 + \nu^+(s))} \Bigg] (1 - x)^{\nu^{+}(s)},	\label{eq-fgp} \\
	f_{g}^-(x,\mu^2) = \Bigg[- {4\over 9} A_{q} e^{- d^{-} s} (1 - x)^{m_g^-} + \\ \nonumber
	+ {K^- \over \Gamma(2 + \nu^-(s))} \times {B^-(s)x \over \hat c - \ln(1 - x) + \Psi(2 + \nu^-(s))}\Bigg](1 - x)^{\nu^{-}(s)+1}.
	\label{eq-fgmtld11}
\end{gather}
\noindent
Here $\Psi(z)$ is the Riemann's $\Psi$-function, $I_0(z)$ and $I_1(z)$ are the modified Bessel functions and
\begin{gather}
	\nu^\pm(s) = \nu^\pm(0) + r^\pm s, \quad B^\pm(s) = B^\pm(0) e^{-p^\pm s}, \quad p^\pm = r^\pm (\gamma_{\rm E} + \hat c^\pm), \nonumber \\
	r^+ = {12\over \beta_0}, \quad r^- = {16\over 3 \beta_0}, \quad \hat c^+ = - {\beta_0 \over 12}, \quad \hat c^- = - {3\over 4}, \quad K^+ = {3N_f \over 10}, \quad K^- = {2\over 5}, \nonumber \\
	\rho = {\sigma \over 2 \ln(1/x)}, \quad \sigma = 2 \sqrt{|\hat d^+|s \ln{1\over x}}, \\ \nonumber
	\hat d^+ = - {12\over \beta_0}, \quad \bar d^+ = 1 + {20 N_f\over 27\beta_0}, \quad d^- = {16 N_f \over 27 \beta_0}
	\label{eq-parameters1}
\end{gather}
\noindent
with $A_g$, $A_q$, $B^\pm(0)$, $\nu^\pm(0)$, $m_q^\pm$, $m_g^\pm$ and $n^\pm$ being free parameters.
The expressions for subasymptotic terms $D^\pm(s)$ were derived from the momentum conservation law
and could be found elsewhere\cite{PDFsSumRules2}.
All the parameters in the formulas above have been determined from a rigorous fit to precision
BCDMS, H1 and ZEUS experimental data on the proton structure function $F_2(x,Q^2)$ in a wide
kinematical region, $2 \cdot 10^{-5} \leq x \leq 0.75$ and $1.2 \leq Q^2 \leq 30000$~GeV$^2$\cite{PDFsSumRules2}.
A perfect goodness of the fit\footnote{The fit was comprising a total of $933$ points from $5$ data sets.}, $\chi^2/n.d.f. = 1.408$, was achieved.

Our next step is connected with nuclear modifications of derived PDFs.
The combination of the rescaling model extended to low $x$ (or shadowing region) and Fermi motion was applied\cite{RescalingModel-Fermi}.
So, for a nucleus $A$, the valence and nonsinglet parts are modified as:
\begin{gather}
	f_{i}^A(x,\mu^2) = f_{i}(x,\mu^2_{A,\,i}),
	\label{va.1a}
\end{gather}
\noindent
where $i = V$ or $NS$ and scale $\mu^2_{A,\,i}$ is related to $\mu^2$ by:
\begin{gather}
	s^A_i \equiv \ln \left(\frac{\ln\left(\mu^2_{A,\,i}/\Lambda^2_{\rm QCD}\right)}{\ln\left(\mu^2_{0}/\Lambda^2_{\rm QCD}\right)}\right) = s +\ln \left(1+\delta^A_i \right),
	\label{sA}
\end{gather}
\noindent
with $\delta^A_i$ being the scale independent free parameters (see\cite{RescalingModel-Apps1} and references therein)
and analytical expressions for $f_i(x,\mu^2)$ are given by~(\ref{eq1-NSV}) --- (\ref{eq3-NSV}).
For singlet quark and gluon parts
which have two ('$+$' and '$-$') independent components,
one has two additional rescaling parameters $\delta^A_\pm$ (see\cite{RescalingModel-Apps1}). So,
\begin{gather}
	f_i^A(x,\mu^2) = f_i^{A,\,+}(x,\mu^2) + f_i^{A,\,-}(x,\mu^2), \quad f_i^{A,\,\pm}(x,\mu^2) = f_i^{\pm}(x,\mu^2_{A,\,\pm}),
\end{gather}
\noindent
where $i = SI$ or $g$ and expressions for $f_i^\pm(x,\mu^2)$ are given by~(\ref{eq1-Sg}) --- (9). 
The definition of $\mu^2_{A,\,\pm}$ is the same as above and
corresponding values of $s^A_\pm$ turned out to be\cite{RescalingModel-Apps1}
\begin{gather}
	s^A_\pm \equiv  \ln \left(\frac{\ln\left(\mu^2_{A,\,\pm}/\Lambda^2_{\rm QCD}\right)}{\ln\left(\mu^2_{0}/\Lambda^2_{\rm QCD}\right)}\right) = s +\ln \left(1+\delta^A_\pm \right).
\end{gather}
\noindent
where the free parameters $\delta^A_\pm$ are scale independent and have to be negative. 

The full model\cite{RescalingModel-Fermi} takes into account the effects of Fermi motion
of nucleons inside the nuclear target. These effects
deform the nuclear structure functions mainly at large $x > 0.7$.
Such deformation could be described by the convolution\cite{FermiMotion1, FermiMotion2}:
\begin{gather}
	f_i^{A(F)}(x,\mu^2)=\frac{1}{R_{i}} \int \frac{dy}{y} \left[yf_N(y)\right] f_i^A\left(x/y,\mu^2\right),
	\label{Fermi1}
\end{gather}
\noindent
where $i = V$, $NS$, $SI$ or $g$.
The expressions for $f_N(y)$ and $R_i$ can be found elsewhere\cite{RescalingModel-Fermi}.
The rescaling parameters $\delta^A_{V}$, $\delta^A_{NS}$ and $\delta^A_{\pm}$
have been extracted\cite{RescalingModel-Fermi} from the global fit on experimental data on structure function ratios
$F_2^A(x,Q^2)/F_2^{A^\prime}(x,Q^2)$
for various nuclear targets $A$ and $A^\prime$ in the deep inelastic region.
Moreover, their nuclear dependence was investigated and
parametrized in several ways, which are often assumed in the literature (see, for example,\cite{QsADependence1, QsADependence2, QsADependence3, QsADependence4, QsADependence5, nCTEQ, ESegarra21, MKla24} and references therein): $\delta^A_i \sim A^{1/3}$ (Fit A),
$\delta^A_i \sim \ln A$ (Fit B) and $\delta^A_i \sim A^{1/3} + A^{-1/3}$ (Fit C).
Thus, within the model\cite{RescalingModel-Fermi}, the nuclear medium modification factor $R$ can be calculated for any nucleus,
even for unmeasured ones.

Now we are able to calculate the nTMDs using the approach\cite{RescalingModel-Fermi} and LO KMR/WMR formalism.
There are known differential and integral definitions of the latter (see discussions\cite{KMR-discussion-1, KMR-discussion-2, KMR-discussion-3, KMR-discussion-4}).
Here we employ the integral definition, which is more preferable\footnote{The analytical expressions for TMDs were obtained\cite{PDFsSumRules2} in both differential and integral definitions of the KMR/WMR approach.}
in the phenomenological applications\cite{KMR-discussion-1}.
In this way, the nTMDs could be written as
\begin{gather}
  f_a^{A}(x, {\mathbf k}_T^2, \mu^2) = T_a({\mathbf k}_T^2, \mu^2) { {\alpha}_s({\mathbf k}_T^2) \over 2\pi {\mathbf k}_T^2} \sum_b \int\limits_x^{1 - \Delta} dz\, P_{ab}(z) f_b^{A(F)}\left({x\over z}, {\mathbf k}_T^2\right),
  \label{eq-KMR-int}
\end{gather}
\noindent
where
$a = q$ or $g$, $P_{ab}(z)$ are the usual unregulated LO DGLAP splitting functions and $\Delta$ is some cutoff parameter.
Usually it has one of two physically motivated forms, $\Delta = \Delta_1 = |{\mathbf k}_T|/\mu$ and $\Delta = \Delta_2 = |{\mathbf k}_T|/(|{\mathbf k}_T| + \mu)$,
which reflects the strong ordering (SO) or angular ordering (AO) conditions for parton emissions
at the last evolution step.
The Sudakov form factors $T_a({\mathbf k}_T^2, \mu^2)$ give the probability of evolving from a
scale ${\mathbf k}_T^2$ to a scale $\mu^2$ without parton emission:
\begin{gather}
	\ln T_a({\mathbf k}_T^2, \mu^2) = - \int\limits_{{\mathbf k}_T^2}^{\mu^2} {d {\mathbf p}_T^2 \over {\mathbf p}_T^2} { {\alpha}_s({\mathbf p}_T^2) \over 2 \pi}
	\sum_b \int\limits_0^{1 - \Delta} dz z P_{ba}(z).
	\label{eq-sudakov}
\end{gather}
\noindent
The analytical expressions for $T_a({\mathbf k}_T^2, \mu^2)$ can be found in\cite{TMD-Laplace}.
It is important to note that the KMR/WMR prescription is
only correct at ${\mathbf k}_T^2 > \mu_0^2$, where $\mu_0^2 \sim 1$~GeV$^2$ is the minimum scale where
the perturbative QCD is still applicable. At low ${\mathbf k}_T^2$ special model assumptions, usually related with
the overall normalization of the TMDs/nTMDs, are necessary (see\cite{KMR-LO, WMR-LO, KMR-discussion-1, KMR-discussion-4}).
However, in our consideration, these parton densities are well defined
in the whole ${\mathbf k}_T^2$ range
since "frozen" treatment of the QCD coupling in the infrared region is applied,
where $\alpha_s(\mu^2) \to \alpha_s(\mu^2 + M_\rho^2)$
with $M_\rho \sim 1$~GeV. Such treatment was used in all the fits\cite{PDFsSumRules2, RescalingModel-Fermi} and results in a good description of the
data on structure function $F_2(x,Q^2)$ and $F_2^A(x,Q^2)$, or rather their ratios.

\begin{figure}
	\begin{center}
		\includegraphics[width=4.75cm]{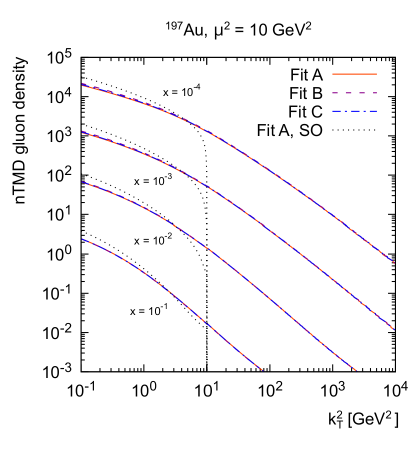}
		\includegraphics[width=4.75cm]{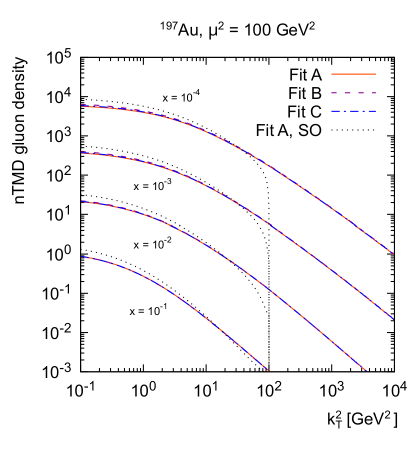}
		\includegraphics[width=4.75cm]{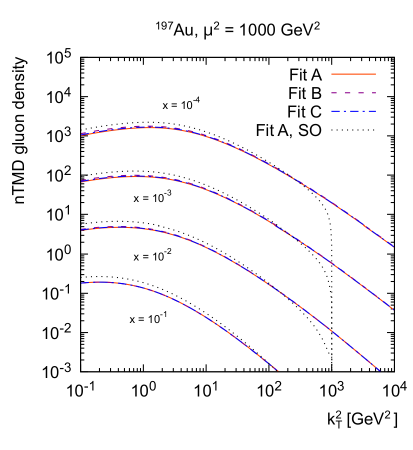}
		\includegraphics[width=4.75cm]{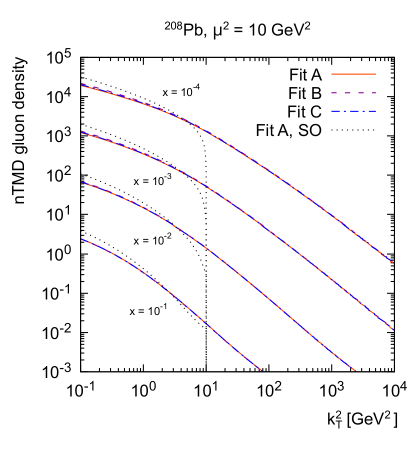}
        \includegraphics[width=4.75cm]{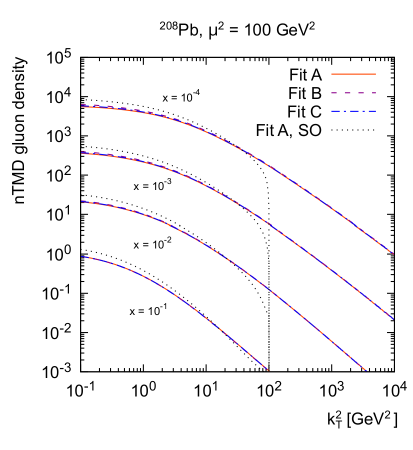}
        \includegraphics[width=4.75cm]{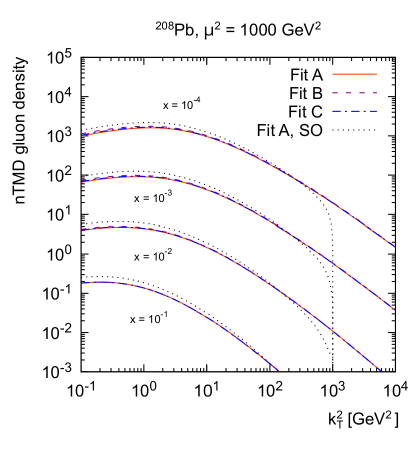}
		\caption{nTMD gluon densities in $^{197}{\rm Au}$ and $^{208}{\rm Pb}$ calculated as functions of ${\mathbf{k}_T^2}$
			for several $x$ and $\mu^2$ using the angular ordering and strong ordering conditions.
			Different scenarios for their nuclear dependence, namely, Fit A, B and C, described in the text (see also\cite{RescalingModel-Fermi}), are applied.}
		\label{fig:1}
	\end{center}
\end{figure}

\begin{figure}
    \begin{center}
        \includegraphics[width=4.75cm]{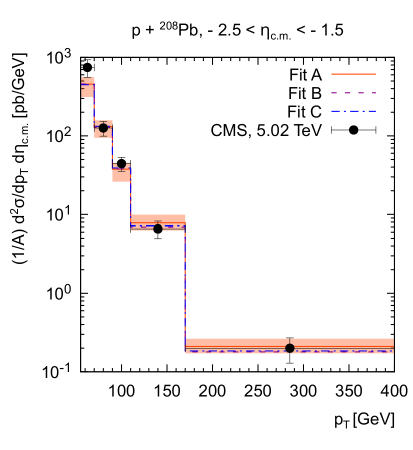}
        \includegraphics[width=6.25cm]{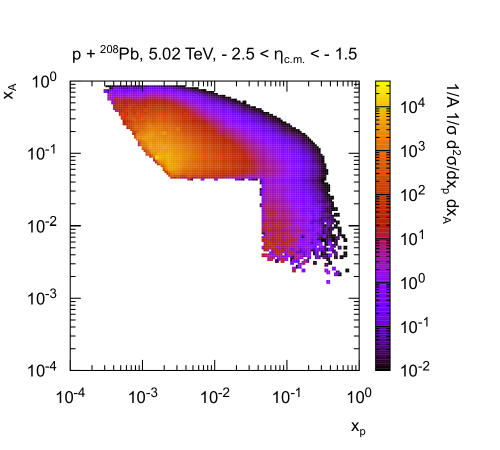}
        \includegraphics[width=4.75cm]{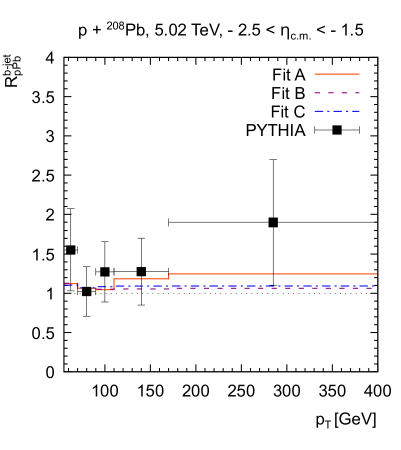}
        \includegraphics[width=4.75cm]{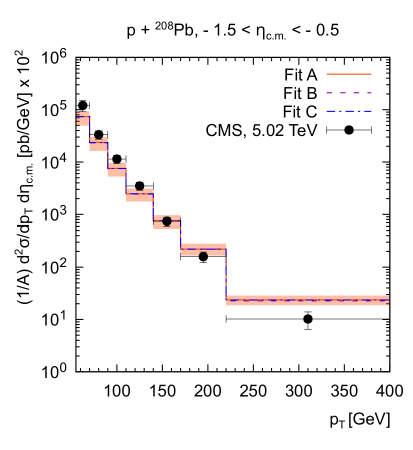}
        \includegraphics[width=6.25cm]{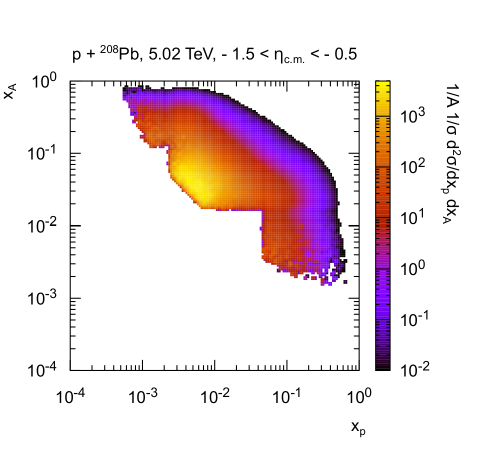}
        \includegraphics[width=4.75cm]{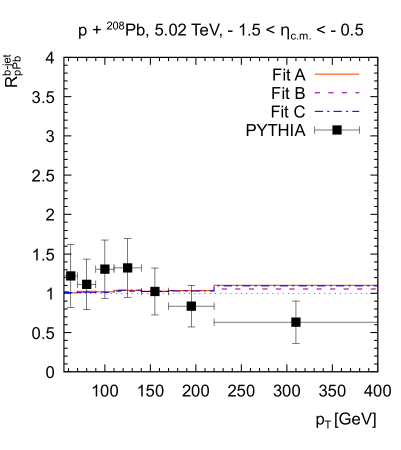}
        \includegraphics[width=4.75cm]{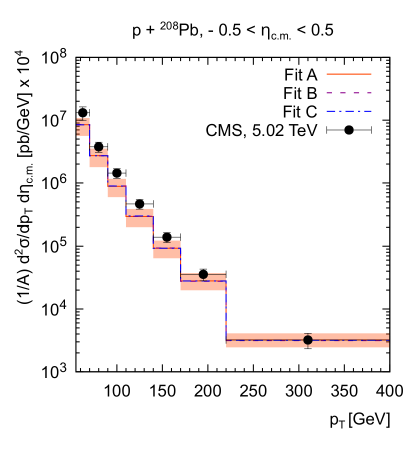}
        \includegraphics[width=6.25cm]{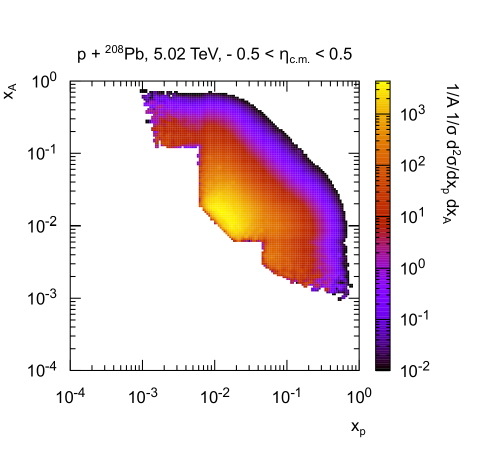}
        \includegraphics[width=4.75cm]{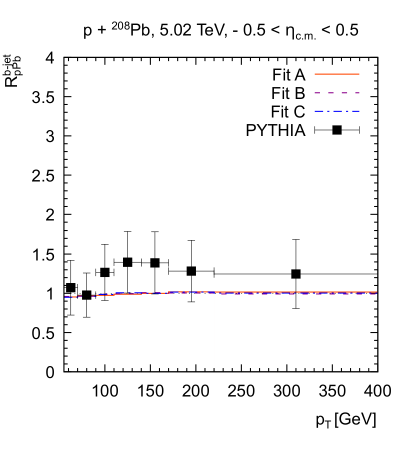}
        \includegraphics[width=4.75cm]{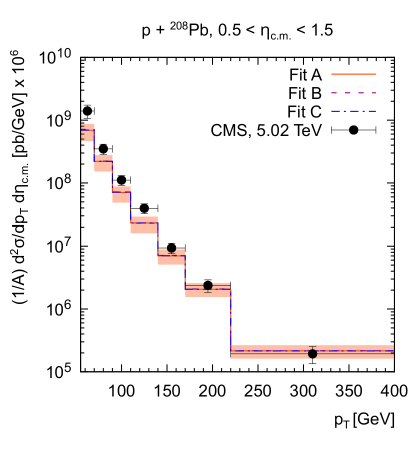}
        \includegraphics[width=6.25cm]{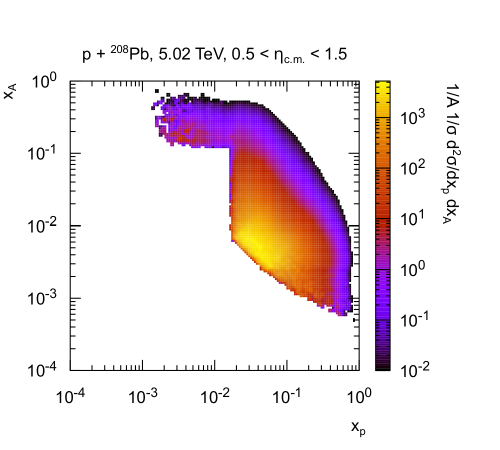}
        \includegraphics[width=4.75cm]{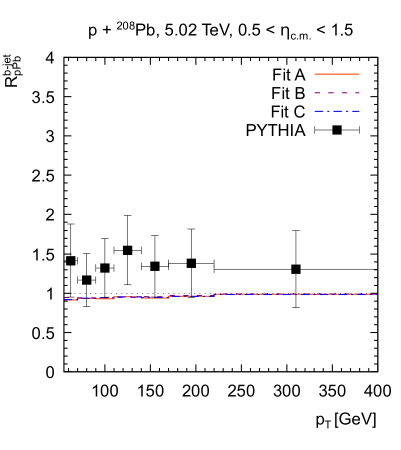}
        \caption{Transverse momentum spectra of inclusive $b$-jet production
compared to CMS data\cite{CMS-bjets} (left panels), normalized
double differential cross sections
as functions of longitudinal momentum fractions of proton, $x_p$, and nucleus, $x_A$ (center panels),
and
nuclear modification factors $R_{p\rm Pb}^{{b\textrm{-jet}}}$ (right panels).
Different scenarios for their nuclear dependence, namely, Fit A, B and C, described in the text (see also\cite{RescalingModel-Fermi}), with AO condition are applied. The uncertainty band is shown for Fit A results only. }
        \label{fig:2}
    \end{center}
\end{figure}

In Fig.~\ref{fig:1} we show the nTMD gluon distributions in $^{197}{\rm Au}$ and $^{208}{\rm Pb}$
calculated as functions of ${\mathbf{k}_T^2}$
for several values of $x$ and scale $\mu^2$
using mentioned above scenarios for nuclear dependence, Fit A, B and C (see also\cite{RescalingModel-Fermi}).
These results were obtained with the angular and strong ordering conditions.
As it was expected, the nTMDs extend
into the ${\mathbf k}_T^2 > \mu^2$ region in the case
of angular ordering and vanish at large ${\mathbf k}_T^2$
in the case of strong ordering condition.
We find that different assumptions on the $A$-dependence
of rescaling parameters
have only weak impact on predicted nTMDs. In fact, results of calculations based on Fit A, B and C
are very close to each other in a wide kinematical range.

As it has already been noted above, here we will test the obtained nTMDs with the data on beauty
production in proton-lead collisions at the LHC.
We strictly follow the previous evaluations\cite{nTMDs-KMR, HF-JKLZ}
done in the High Energy Factorization\cite{HighEnergyFactorization}, or $k_T$-factorization\cite{kt-factorization} approach
and employ the FFNS scheme with $N_f = 4$.
In such a way, the main contribution to beauty production
comes from the off-shell
gluon-gluon fusion subprocess $g^* + g^* \to b + \bar b$, which is
implemented into the Monte-Carlo event generator \textsc{pegasus}\cite{PEGASUS}.
The contribution from the quark-induced processed is of almost no
importance because of comparatively low quark densities.
The cross section is given by
the convolution of the TMD gluon
densities in a proton and/or nuclei and off-shell (dependent on the transverse momenta of incoming partons) hard scattering amplitudes (see\cite{TMD-review, TMD-review-our} for more information).
Note that MC generator \textsc{pegasus}
was used, in particular, by the ATLAS\cite{PEGASUS-ATLAS1, PEGASUS-ATLAS2} and ALICE\cite{PEGASUS-ALICE1, PEGASUS-ALICE2} Collaborations in the analyses of their data on heavy quark bound states production at the LHC.

\begin{figure}
    \begin{center}
        \includegraphics[width=4.75cm]{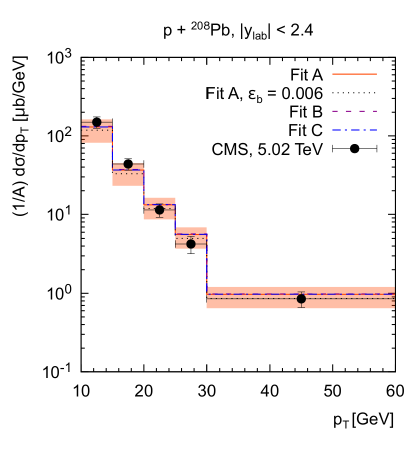}
        \includegraphics[width=4.75cm]{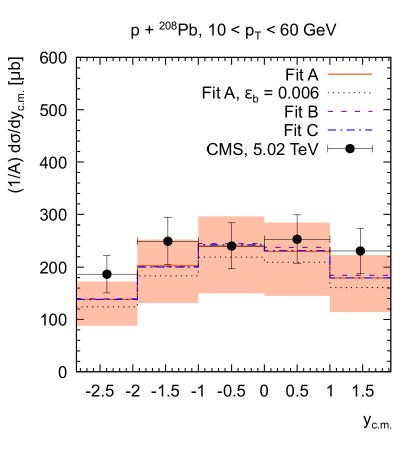}
        \includegraphics[width=4.75cm]{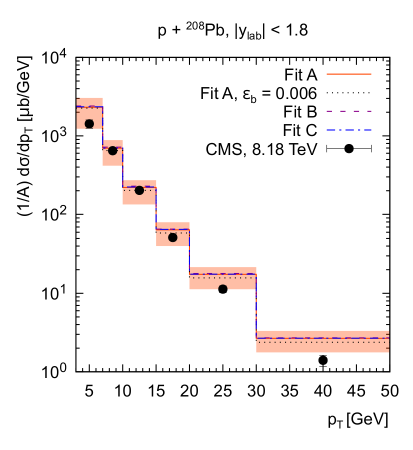}
        \includegraphics[width=4.75cm]{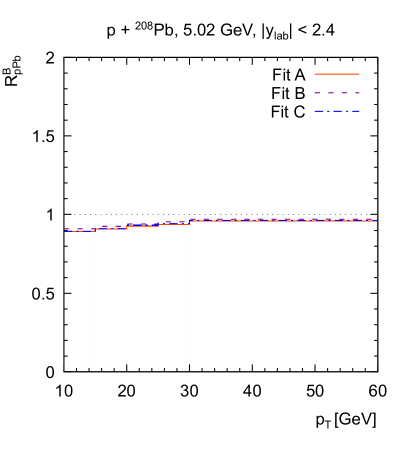}
        \includegraphics[width=4.75cm]{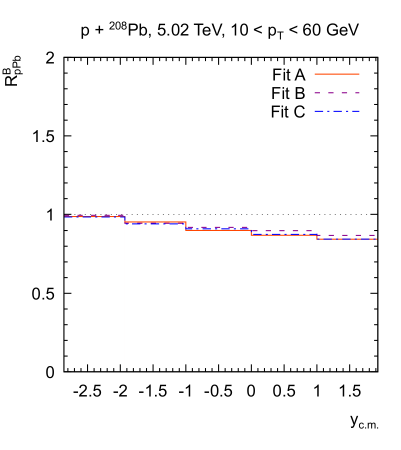}
        \includegraphics[width=4.75cm]{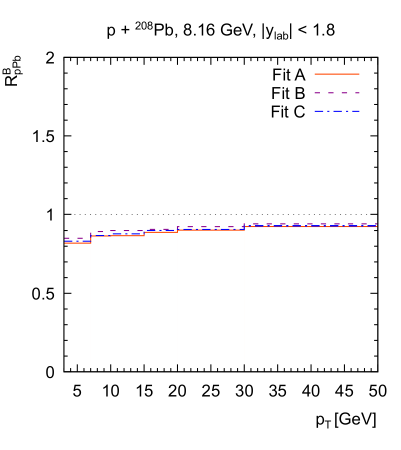}
        \caption{Transverse momentum and rapidity spectra of inclusive $B^+$-jet production
compared to CMS data\cite{CMS-B,CMS-B8} (upper panels)
and corresponding
nuclear modification factors $R_{p\rm Pb}^{B^+}$ (lower panels). Different scenarios for their nuclear dependence, namely, Fit A, B and C, described in the text (see also\cite{RescalingModel-Fermi}), with AO condition are applied. The uncertainty band is shown for Fit A results only.}
        \label{fig:3}
    \end{center}
\end{figure}

We consider the latest measurements of inclusive $b$-jet and $B^+$ meson production performed by the CMS Collaborations at $\sqrt s = 5.02$ and $8.16$~TeV\cite{CMS-bjets, CMS-B, CMS-B8}.
The CMS data on $b$-jet transverse momentum spectra\cite{CMS-bjets} refer to the
transverse momentum range $55 < p_T < 400$~GeV and
were obtained at $\sqrt s = 5.02$~TeV in four different proton-lead center-of-mass pseudorapidity subdivisions, namely, $-2.5 < \eta_{\rm c.m.} < -1.5$,
$-1.5 < \eta_{\rm c.m.} < -0.5$, $-0.5 < \eta_{\rm c.m.} < 0.5$ and $0.5 < \eta_{\rm c.m.} < 1.5$.
Note that the positive rapidities refer to the beam orientation where the
proton is moving toward positive $z$ direction.
The transverse momentum and rapidity distributions of $B^+$ mesons have been measured at $10 < p_T < 60$~GeV, $|y_{\rm lab}| < 2.4$
and $\sqrt s = 5.02$~TeV\cite{CMS-B}. Very recently, first data on $B^+$ meson production at $\sqrt s = 8.16$~TeV have been
reported in the kinematical range $3 < p_T < 50$~GeV and $|y_{\rm lab}| < 1.8$\cite{CMS-B8}.
In the present calculations, we identify the produced $b$ quark with the beauty jet.
To convert $b$ quarks to $B^+$ mesons, we apply standard Peterson fragmentation function with 
shape parameter $\epsilon_b = 0.003$, which is often used in the pQCD calculations. We
set $m(b) = 4.183$~GeV, $m(B^+) = 5.279$~GeV and branching ratios $B(b \to B^+) = 0.408$\cite{PDG}.
Following to fits\cite{RescalingModel-Fermi, PDFsSumRules2}, the calculations were made with the one-loop formula
for QCD coupling $\alpha_s$ with $\Lambda_{\rm QCD}^{(4)} = 118$~MeV.

Our predictions are shown in Fig.~{\ref{fig:2}} and Fig.~{\ref{fig:3}}.
In Fig.~{\ref{fig:2}} we plot the
transverse momentum spectra of inclusive $b$-jet production
compared to the CMS data\cite{CMS-bjets}, normalized
double differential cross sections
as functions of longitudinal momentum fractions $x_p$ and $x_A$
(carried by the gluons originated from
proton and nuclear target, respectively)
and corresponding
nuclear modification factors $R_{p\rm Pb}^{{b\textrm{-jet}}} = 1/A (d\sigma^{{b\textrm{-jet}}}_{p \rm Pb}/dp_T)/(d\sigma_{pp}^{{b\textrm{-jet}}}/dp_T)$,
where $A = 208$ is the number of nucleons in the Pb nucleus.
The results of calculations for $B^+$ meson production are shown in Fig.~{\ref{fig:3}}.
The AO condition is applied everywhere and calculations are done in the each of pseudorapidity subdivisions
considered in the CMS analysis.
Solid, dashed and dash-dotted histograms corresponds to the
central results for different scenarios of nuclear dependence,
where we fixed both renormalization and factorization scales at their default values,
$\mu_R^2 = \mu_F^2 = p_T^2$.
The shaded bands correspond to scale uncertainties of the calculations,
shown for Fit A scenario.
These uncertainties have been estimated in a standard way, by varying the scales $\mu_R$ and $\mu_F$
by a factor of $2$ around their default values.
We find that our predictions
based on Fit A, Fit B and Fit C parametrizations of nuclear dependence
are rather close to each other
and consistent with the data
within the theoretical and experimental uncertainties. However,
overall agreement achieved in rear kinematical region (negative $\eta_{\rm c.m.}$) is somewhat better
than in the forward one.
In fact, our results
tend to slightly underestimate the CMS data at forward rapidities.
It is clear that main contribution in this region
comes from events where essentially small-$x_A$ region of nuclear target $A$ is probed (see Fig.~{\ref{fig:2}}).
Thus, the observed difference 
may indicate that the
nuclear shadowing effects, which manifest themselves mainly at low $x_A$,
could be overestimated in our analysis.
Another point is related with
effects of parton showers and/or hadronization in final state.
However, it is difficult to give a definite conclusion
due to relatively large theoretical uncertainties of our calculations
and shortage of the nuclear experimental data used in the determination of rescaling parameters (see also\cite{RescalingModel-Fermi}).
Moreover, additional higher-order pQCD
contributions (not covered by the non-collinear pQCD evolution encoded in the TMDs/nTMDs)
also could play a role (see, for example,\cite{chicNLO-kt, psiNLO-kt} and references therein).
An accurate addressing to all of these issues needs a dedicated study and is out of the present consideration.

It was already pointed out that special
kinematical cuts on the final state provide possibilities to
achieve the wanted region of $x_A$\cite{MapForTMD}.
So, with increasing $\eta_{\rm c.m.}$ the peak of the normalized
cross section $1/\sigma \, d^2\sigma/dx_p dx_A$
shifts gradually to lower $x_A$ values,
which immediately reflects in the predicted
nuclear medium modification factor $R_{p\rm Pb}^{{b\textrm{-jet}}}$ (see Fig.~{\ref{fig:2}}).
In particular, our calculations show
sizeble antishadowing and shadowing effects
with $R_{p\rm Pb}^{{b\textrm{-jet}}} \sim 1.1 - 1.2$
and $R_{p\rm Pb}^{{b\textrm{-jet}}} \sim 0.9 - 1.0$
in the rear and forward kinematical regions, respectively.
Similar values are obtained for $B^+$ mesons (see Fig.~{\ref{fig:3}}).
Thus, we demonstrate that one can investigate the nuclear medium effects by applying the
different cuts on the final states produced in the high energy $pA$ collisions.
This is important to precise determination of the
partonic content of nuclei, where there are a lot of differences among predictions
from different groups\footnote{The pseudorapidity-integrated value predicted by the \textsc{pythia} event
generator\cite{PYTHIA6.4} is
$R_{p\rm Pb}^{{b\textrm{-jet}}} = 1.22 \pm 0.15$~(stat. + syst. exp.)~$\pm 0.27$~(syst. \textsc{pythia}), see\cite{CMS-bjets}. To compare with our results, in Fig.~\ref{fig:2} we show the \textsc{pythia}
predictions (taken from\cite{CMS-bjets}) separately for each rapidity subdivision.}.

Finally, to investigate the dependence of our predictions on the $b \to B^+$
fragmentation function, we repeated the calculations with the
shifted value of the Peterson shape parameter, $\epsilon_b = 0.006$.
The results of our calculations are shown in Fig.~\ref{fig:3} by dotted histograms.
We find that the obtained cross sections are somewhat smaller for larger $\epsilon_b$ values.
However, these predictions lie within the estimated theoretical uncertainties.

To conclude,
we have evaluated nTMDs at the leading order in QCD coupling
using the popular Kimber-Martin-Ryskin/Watt-Martin-Ryskin approach.
The calculations have been based on a rescaling model
with taking into account Fermi motion effects
and global analysis of available DIS
data for different nuclear targets.
The obtained nTMDs have been successfully tested with latest CMS data on
beauty production in proton-lead collisions collected at $\sqrt s = 5.02$ and $8.16$~TeV.
We predict the corresponding nuclear medium modification factors
to be about of $0.8 - 1.2$.
Specially we have highlighted the possibility to investigate the
nuclear modification of parton densities by applying
different cuts on the final states produced in the $pA$ collisions.
It can be important for forthcoming studies of lepton-nucleus, proton-nucleus and
nucleus-nucleus interactions at modern and future colliders, where nuclear
parton dynamics could be examined directly.

Note that the consideration
can be improved by taking into account
effects of parton showers and/or higher-order pQCD corrections.
We plan to investigate these issues in ensuing studies.


{\sl Acknowledgements.} We thank S.P.~Baranov, M.A.~Malyshev and H.~Jung for their
interest, very important comments and remarks. We are also grateful to A.Yu.~Alekseenko for testing the TMDs used in this work.
This research has been carried out at the expense of the Russian Science Foundation grant No.~25-22-00066, https://rscf.ru/en/project/25-22-00066/.


\bibliography{nTMD-1}

\end{document}